\documentclass[prd,twocolumn,aps,superscriptaddress,eqsecnum,floatfix]{revtex4-1}
\usepackage{empheq}
\usepackage{amsmath}
\usepackage{amssymb}
\usepackage{dsfont}
\usepackage{graphicx}
\usepackage{hyperref}
\usepackage{stackrel}
\usepackage{color}
\usepackage{comment}

\def \ee{\end{equation}}
\def \be{\begin{equation}}
\def \eea{\end{eqnarray}}
\def \bea{\begin{eqnarray}}

\begin{document}

\title{Electromagnetic stress on nucleon structure}

\author{Benjamin Koch}
\email{bkoch@fis.puc.cl}
\affiliation{Pontificia Universidad Cat\'olica de Chile \\ Instituto de F\'isica, Pontificia Universidad Cat\'olica de Chile, \\
Casilla 306, Santiago, Chile}
\affiliation{Institut f\"ur Theoretische Physik,
 Technische Universit\"at Wien,
 Wiedner Hauptstrasse 8-10,
 A-1040 Vienna, Austria}

\begin{abstract}
External electromagnetic fields can provoke stress, and thus modifications of the internal structure of nucleons. 
Working with this hypothesis, one can
 derive a simple 
description of the charge dependence of the EMC effect. 
This first result is confirmed by two
 explicit models of the structure functions
of deformed nucleons in the atomic nucleus.
For  large nuclei a continuous model is used.
For small nuclei a discrete distribution of nuclear matter gives better results.
\end{abstract}

{\color{blue}
}
\maketitle


\tableofcontents

\section{Introduction}

Three forces influence the composition, structure, and stability of protons and neutrons (nucleons): 
The strong force,
binding the nucleus together,
the weak force, hold responsible for rare weak decays,
and the electromagnetic force, typically playing a secondary role.
Nucleons contain different types of particles; in particular, quarks
carry charges of all three forces, and gluons propagate the strong force only.

For the following discussion, the important fact is that
the quarks carry both electrical and color charges.
Even though the electrical charges of some quarks repel each other, 
the nucleon is held
together by the dominant forces of the gluon field $\sim F_{int}$.
However, if one would expose this system for example to an external electric field ${ F}_{ext}$,
the quarks would feel this effect directly, while the gluons would feel it only indirectly, through strong interactions
with those quarks.

A difference in responding to external electromagnetic forces will 
necessarily tend to change the spatial composition and even form
of the nucleon. 
This mechanism will be called and abbreviated as Electro-Magnetic Stress (EMS).
An estimate of this EMS effect can be obtained
if one considers the interplay between the positive and the negative electrically charged constituents of a nucleon.
Those constituents will be polarized in external electromagnetic fields, which is
why this effect is known as ``hadron polarizability''.
The corresponding hadronic dipole moment is
\be
\vec p = 4 \pi \alpha_E \vec E.
\ee
For a review on this topic, see~\cite{Holstein:2013kia}.
In experiments at very low $\mathcal{Q}^2$ and with isolated nucleons,
the proportionality factor $\alpha_E$ (not to be confused with the
electromagnetic coupling $\alpha_e$) has been
measured to be of the order of $\alpha_E \approx 10^{-3} fm^3$~\cite{Holstein:2013kia}.
Between two neighboring protons, this would correspond
to a relative deformation of the proton radius of
\be\label{delta1}
\delta = \frac{\Delta r}{r}\sim \frac{\alpha_E}{r_p^3}\approx 0.001.
\ee
Please check the appendix for a derivation of this estimate.

In this article, the testability of
changes in the nucleon structure caused by
external forces like $F_{ext}\approx q E_{ext}$ will be explored in the context of the EMC effect.
In the following section
\ref{sec:DIS},
deep inelastic scattering will be discussed
and a novel application to the  
puzzling EMC effect (named after the
``European Muon Collaboration'') will be presented.
This argument will be consolidated in section \ref{sec:model},
where the effect is shown to work with an explicit model for nuclear structure functions.
Finally, in section \ref{sec:sum}, a  summary of 
the results is given.

\section{Expected proportionality 
in deep inelastic scattering}\label{sec:DIS}

A manifestation of the electromagnetic stress on the
nucleon structure can be found in the context of Deep Inelastic Scattering (DIS).

\subsection{General discussion}
The working hypothesis is that the neutron and proton structure is sensitive 
to the force $\bar F$ produced by external electromagnetic fields in the rest frame.
If the structure gets modified in one frame,
it will get modified in any other frame as well.
The structure functions like $F_2$, are a weighted sum 
over the parton content $f_q$.
In the presence of an average external electric $\bar F_E\sim \bar E q$ 
and magnetic $\bar F_B\sim \nabla (\mu {\bar B})$ forces 
these structure functions will get modified
\be
F_2=F_2(x,\bar F_E, \bar F_B,s)\approx F_2(x,0) \left( 1+ (\bar F_E+g \bar F_B+s) d(x)\right).
\ee 
Here, $x$ is the fraction of the nucleon momentum contributed by a parton in a collision.
Please note that the parton model itself is defined for a highly boosted 
light cone frame but the dependence on external electromagnetic fields is for convenience
calculated in the rest frame of the nucleon.
Lorentz contraction might lead to different modifications in different directions
when going from one frame to another.
Thus one has to take the force averaged over the nucleus $\bar F$ instead of 
the local directed force $\vec F$.
The correction due to the external fields $\bar F_E, \bar F_B$ is proportional to $d(x)$
which for small variations around a small $x_0$ can be approximated to $d(x)\approx d\cdot  (x-x_0)$.
The constant $s$ summarizes other short range and surface interactions.
Here, $d,g$, and $x_0$ are the proportionality constants in this linear expansion
 that will have to be determined experimentally.
At leading order, this modification will be reflected one to one 
if one compares the deep inelastic  $x$-dependent cross-sections of a neutron (proton) with and without 
external electromagnetic fields and surface interactions. One finds
\bea\label{fxexp}
\frac{\sigma(x)_{\bar F_E,\bar F_B}}{\sigma(x)_{0}}
&\sim& \frac{F_2(x,\bar F_E, \bar F_B,s)}{F_2(x,0)}\\ \nonumber
&\approx&  \left( 1+
 (\bar F_E+ g \bar F_B+s) d \cdot (x-x_0)\right),
\eea
where $g,s,d,x_0$ are constants.
The first approximate proportionality in equation (\ref{fxexp})
comes from the fact that, at leading order in $m_p^2/{\mathcal{Q}}^2$, 
the corresponding differential cross-section 
is proportional to the structure function~$F_2$~\cite{DeRoeck:2011na}.
The $x$ derivative of the ratio (\ref{fxexp}) is
\be 
\frac{d}{dx}\frac{\sigma(x)_{\bar F_E, \bar F_B}}{\sigma(x)_{0}}\approx  \bar F_E d+\bar F_B gd+ds.
\ee

\subsection{EMC effect}

In the most simple version of an atomic nucleus model, one would expect
that  the DIS cross-section is given by  the cross-section of
a nucleon
multiplied by the number of participants $A$ (neutrons+protons) in this collision
\be\label{simple1}
\sigma^{A,Z}(x)=\frac{A}{2} \sigma^{2,1}(x).
\ee
Thus, the DIS data for heavy nuclei should be predictable from the
DIS data obtained from lighter nuclei and vice versa. In particular, the ratio
\be\label{EMCratio0}
\Delta^{A,Z}_{EMC} =\frac{2\sigma^{A,Z}(x)}{A\cdot \sigma^{2,1}(x)}
\ee
was expected to be constant (one) and not a function of Bjorken $x$.
The EMC effect
is the famous observation that $\Delta^{A,Z}_{EMC}=\Delta_{EMC}^{A,Z}(x)$~\cite{Aubert:1983xm}.
This observation has triggered numerous experimental tests and theoretical explanations.
An incomplete list includes for example nuclear binding \cite{Akulinichev:1985ij,Akulinichev:1986gt,Dunne:1985cn,Dunne:1985ks,Bickerstaff:1989ch,Benhar:1997vy}, pion excess \cite{Ericson:1983um,Wiringa:1984tg,Berger:1984na,Berger:1987er}, multi-quark clustering \cite{Jaffe:1982rr,Carlson:1983fs,Chemtob:1983zj,Clark:1985qu},
dynamical rescaling \cite{Nachtmann:1983py,Close:1983tn}, medium modification \cite{Bentz:2001vc,Mineo:2003vc,Smith:2003hu,Cloet:2006bq,Cloet:2009qs,Cloet:2012td} and short-range correlations \cite{CiofiDegliAtti:1989eg,Weinstein:2010rt,Fomin:2011ng,Arrington:2012ax,Hen:2012fm,Frankfurt:2012qs,Hen:2013oha}.
For a review see, for example  \cite{Malace:2014uea,Geesaman:1995yd}.

Since the shadowing and anti-shadowing effects of the EMC effect at  low $x<0.2$
are well understood, we will first focus on the linear intermediate  regime $0.2<x<0.6$.
In particular, it was noted that
the size of the EMC effect depends on the charge $Q$ of the atomic nuclei~\cite{Gomez:1993ri}.
Below, this effect will be studied in the light of our hypothesis of electromagnetic stress acting on nucleons,
a perspective which we have not found in the literature.\\

\begin{figure}[h!]
\includegraphics[width=\columnwidth]{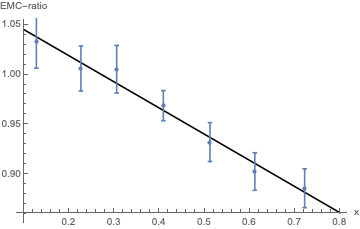}
\caption{
$x$-dependence of the ratio (\ref{EMCratio}) for the case of aluminium \cite{Gomez:1993ri}, allowing 
for a linear fit with $\Delta^{27,13}_{EMC}(x)=(1.07\pm 0.01)-(0.26\pm 0.02)\cdot x$,
where the given errors are the standard statistical errors of the fit.
The fit stops below $0.8$, because at higher $x$ the data ceases to be linear and
higher orders in $(x-x_0)$ would be necessary fitting this behavior.
}
\label{fig:Al}
\end{figure}

Let's return to the most simple version of an atomic nucleus model but allowing for 
contributions of electromagnetic stress.
Larger atomic nuclei will have larger charge $Q$  and thus accumulate larger average electric fields $\bar{E}(Z)$
of the surrounding protons. Therefore,
one has to correct the relation (\ref{simple1}) by
\be
\sigma^{A,Z}(x)=\frac{A}{2} \sigma_{\bar F_E(Z)}^{2,1}(x).
\ee
In this straightforward model, one can compare the normalized  DIS $x$-dependent cross-section of large nuclei
with the DIS cross-section of small nuclei
\bea\label{EMCratio}
\Delta^{A,Z}_{\bar F}&=&
\frac{2\sigma^{A,Z}(x)}{A\cdot \sigma^{2,1}(x)}\\ \nonumber
&\approx& \left( 1+ (\bar F_E(Z)+g \bar F_B+s) d \cdot (x-x_0)\right),
\eea
where the expansion point $x_0$ was chosen to be the point for which this ratio
crosses the value of one with a negative slope.
For a given isotope, this linear dependence  reads $\sim 
a'+b'\cdot x$. It can be obtained by fitting
the experimental data for (\ref{EMCratio}), as shown in figure \ref{fig:Al} for the case of an aluminum isotope.

This fit was repeated for the elements $(He,Be, C, Al,Ca,Fe,Ag,Au)$, using  data from
\cite{Gomez:1993ri} and the digitalization tool \cite{WebDigit}.
For other recent findings on nuclear dependence of the EMC effect 
see \cite{Seely:2009gt,Fomin:2011ng,Arrington:2012ax}.
One finds for example that 
$x_0=0.27\pm 0.02$.

Even more interesting information can be obtained from 
 the $x-$slope, because it will allow relating to $\bar F_E=\bar F_E(Z)$.
Deriving (\ref{EMCratio}), with respect to $x$, gives
\be\label{EMCmod}
\frac{d}{dx}
\Delta^{A,Z}_{\bar F}\approx  (\bar F_E(Z)+g \bar F_B+s) d .
\ee
The magnetic field within the nucleus will be produced
by the surrounding spin $1/2$ protons and neutrons. However, those are
oriented randomly 
\cite{Ma:1997gy} and thus, one can expect the $Q$-dependence of $\bar F_B$ in (\ref{EMCmod}) 
to be subdominant in comparison with $\bar F_E$.
The same holds for the short-range contributions $s$.
Let's now estimate
the average electric force as a function of nucleon charge~$\bar F_E\sim q\bar E(Z)$
by assuming a constant average charge density $\rho$.
The total charge of such a spherical nucleus is
\be
Z=\frac{4}{3}\pi R^3 \rho,
\ee
which can be solved for the nucleon radius.
The radial electric field which is produced by the surrounding protons of a nucleon
within the same nucleus is obtained from Gauss law
%
\be\label{Gauss}
 E_r(r)=\frac{1}{\epsilon_0 3}   r\rho .
\ee
The average value of this field is
\bea\nonumber
\bar E&=&\frac{1}{\epsilon_0 3}  \rho \frac{\int_0^R dr'r'^2 \cdot r' }{\int_0^R dr'r'^2}=\frac{1}{\epsilon_0 4}  \rho R\\
&=&\frac{1}{\epsilon_0 4}  \rho \left(\frac{3Z }{4\rho \pi}\right)^{1/3}\sim Z^{1/3}.\label{barE}
\eea
Note that the field  (\ref{barE}), produced
by the neighboring protons, is the most dominant one.
All other contributions, like
the field produced by the eventually surrounding electrons 
or any other external field of the experimental apparatus
can be neglected.

Inserting (\ref{barE}) into (\ref{EMCmod})
one obtains a prediction of the charge dependence of the slope data
\be\label{EMCmod2}
\frac{d}{dx}
\frac{2\sigma^{A,Z}(x)}{A\cdot \sigma^{2,1}(x)}\approx  g'+Z^{1/3} d'.
\ee
The constant $g'$ is expected to originate for example from the
largely $Z$-independent average magnetic
field $\bar F_B$, or other mean-field and boundary effects.
Note that usually the ratio (\ref{EMCmod2}) is plotted as a function 
of atomic number $A$ and not as a function of charge $Z$. However,
this distinction is a subleading effect since
$A$ and $Z$ are proportional within the precision of the slopes given here.

\begin{figure}[h!]
\includegraphics[width=\columnwidth]{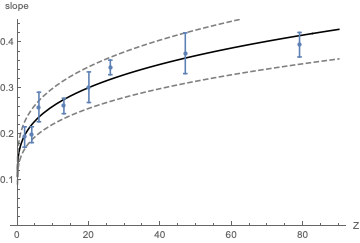}
\caption{
Charge dependence of (\ref{EMCmod2}). The best fit to the experimental data is
obtained for $g'=0.11\pm 0.02$ and $d'=0.072\pm 0.007$.
The data points show the
statistical error bars of the slope-fits of the ratio (\ref{EMCratio}).
The dashed gray lines are an estimate of systematic error of the model (\ref{barE}), 
which is at least $15\%$ 
due to the limited validity of the assumption that 
the charge over mass density $\rho$ is constant for all charges and radii.
}
\label{fig:slopeQ}
\end{figure}

The EMC-type ratio has been measured and fitted like in figure \ref{fig:Al} for eight different atomic nuclei $(He,Be, C, Al,Ca,Fe,Ag,Au)$, which allows extracting the observed data for the predicted charge dependence~(\ref{EMCmod2}).
As shown in figure \ref{fig:slopeQ}, one gets a good
match between (\ref{EMCmod2}) and the data obtained from the slope fitting
like the one plotted in figure \ref{fig:Al}.
Given the simplicity of the underlying idea and model,
the good agreement between the model (\ref{EMCmod2}) 
and the data in figure \ref{fig:slopeQ} is  remarkable.

Note that usually the slope graph \ref{fig:slopeQ} is shown 
as a function of atomic number $A$ and not of atomic
charge, but since both quantities are proportional up to $10\%$,
there is no substantial difference between both plots.

\section{An explicit model of deformed nucleons in the atomic nucleus}\label{sec:model}


\subsection{Distribution of deformations in the nucleus}

\begin{figure}[h!]
\includegraphics[width=2\columnwidth/5]{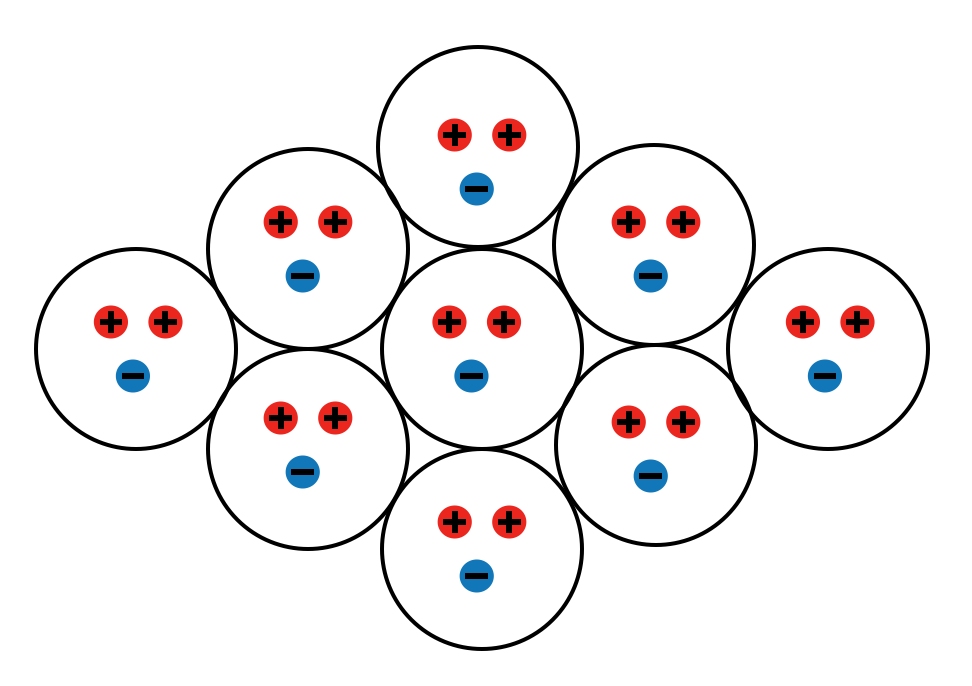}
\includegraphics[width=2\columnwidth/5]{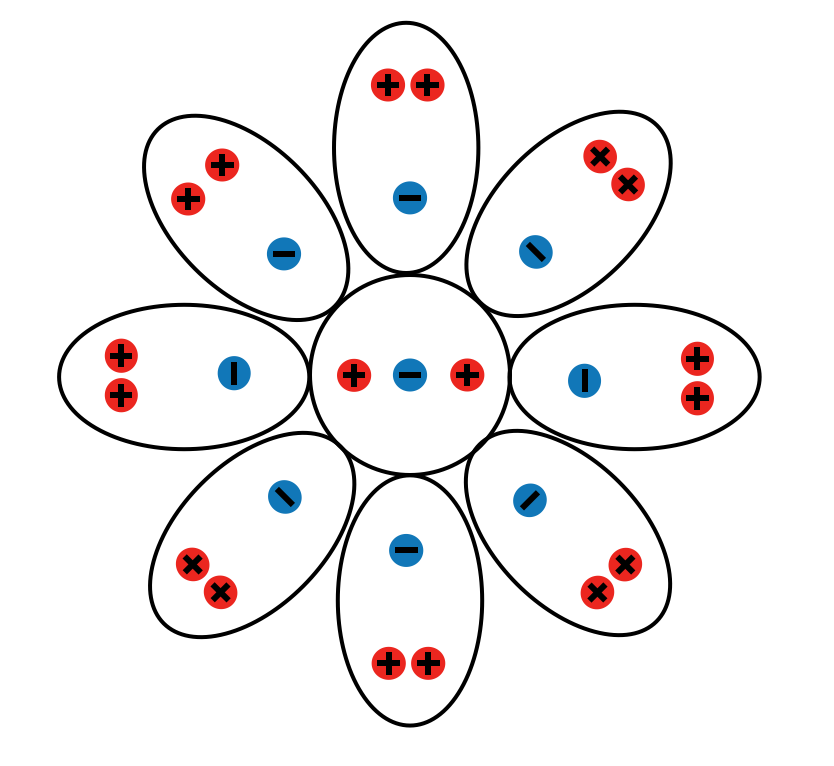}
\caption{
Toy nucleus without (left)
and with (right) EMS}.
\label{fig:NoStress}
\end{figure}

The difference between stress and ``no-stress'' can be easily visualized graphically for a toy example.
In this example, the toy nucleons within a nucleus all have charge one and are composed of 
two partons with charge $+1$, one parton with charge $-1$, and the electrically neutral but
 strongly interacting rest, which is confining the nucleon.
When neglecting the EM stress, all nucleons shall be described by
identical spheres independent of their position within the nucleus,
as shown on the left-hand side of figure \ref{fig:NoStress}.

This changes when one takes into account the stress felt
by the charged partons of an individual nucleon,
which is produced due to the '`external'' electric field of the surrounding
nucleons.
This stress will tend to pull the negative partons towards
the center of the nucleus and push the positive partons towards
the outer border of the nucleus and thus deform the nucleon,
as shown on the right-hand side of figure \ref{fig:NoStress}.
Please note that the toy-structure shown in figure \ref{fig:NoStress} is only chosen for illustrative 
purposes, in order to visualize the EMS mechanism.

The physically relevant question is, 
whether this electromagnetic stress which was estimated in (\ref{delta1}) is strong enough
to be  observable in deep inelastic scattering.
Apart from the pure size of the deformation, there is an additional effect. 
In large nuclei, the amount and direction of this deformation will depend on the position
of the nucleon within the nucleus, as shown on the right side of figure \ref{fig:NoStress}.
This dependence, which goes beyond a mean-field approximation, will be considered now.

\begin{figure}[h!]
\includegraphics[width=\columnwidth*7/8/2]{WithStress.png}
$\quad$
\includegraphics[width=\columnwidth/2]{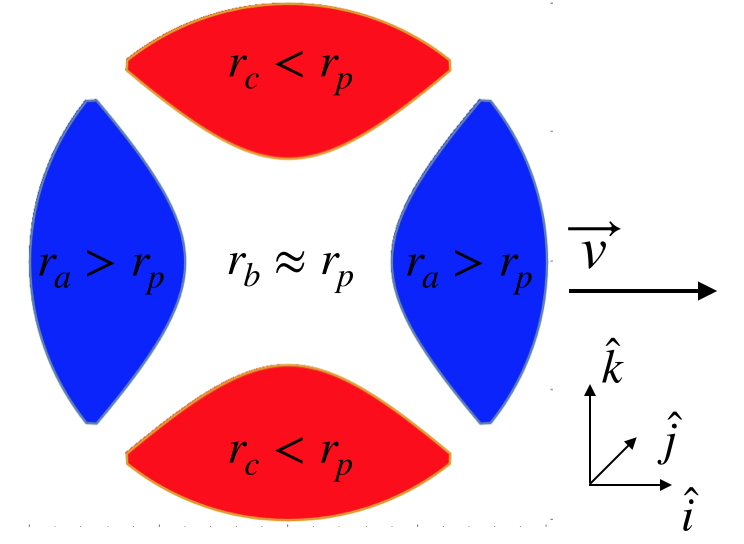}
\caption{\\Left:
Toy nucleus with EM stress.\\
Right: 
Nucleon distribution in an atomic 
nucleus. Nucleons stretched in  $\hat i$-direction are indicated with blue,
nucleons squeezed in $\hat i$-direction with red, and nucleons with negligible deformation
in $\hat i$ with white. When integrating the colored volumes in the right figure,
one finds that red, white, and blue occupy 54\%, 27\%, and 19\% of the volume.
Note that this integration has a rotational symmetry with respect to the $\hat i$ direction.}
\label{fig:WithStress}
\end{figure}

One expects that the EMS generated deformation is directed radially outward
and that its amount increases with
distance from the center of the nucleus.
For example, the electric field of a spherically symmetric distribution
of constant charge density $\rho$ scales as (\ref{Gauss})
\be\nonumber
 E_r(r)\sim   r\rho .
\ee
Concerning this,  one
can distinguish three different regions of deformation within a large nucleus,
as shown in figure~\ref{fig:WithStress}.
\begin{itemize}
\item[a)] In the front and back of the nucleus moving with velocity $\vec v$, the thickness
 in direction $\hat i$ is larger than the radius of an undeformed nucleon $r_a>r_p$.
This is indicated by the blue region on the right-hand side of figure \ref{fig:WithStress}.
\item[b)] Nucleons in the central part of the nucleus
will have negligible stress. The outer nucleons in the diagonal region (e.g. $\sim \pm (\hat k \pm \hat i)$)
will be deformed, but their thickness projected on the $\hat i$ direction will be
approximately the same as the radius of an undeformed nucleon $r_b \approx r_p$. 
Those cases with $r_b \approx r_p$ are indicated
by the white region on the right side of figure \ref{fig:WithStress}.
\item[c)]  Nucleons in the lateral belt  will be thinner in $\hat i$ direction
than undeformed nucleons $r_c< r_p$. This is indicated by the red region on the right side of figure \ref{fig:WithStress}.
\end{itemize}
The fraction of nucleons in each of 
those categories will be labeled $p_a,\, p_b, \, p_c$.
If the scattering particle interacts with a long-range interaction, which means
it sees the entire
nucleus, one expects that $p_a+p_b+p_c=1$. Further,
for a sizable polarization, one expects the three fractions
to be of comparable size, like in figure \ref{fig:WithStress}.


\subsection{Edin model for the nucleon}
\begin{figure}[h!]
\includegraphics[width=\columnwidth]{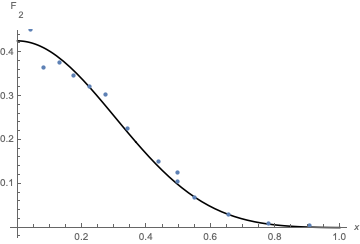}
\caption{Fit of the proton structure function $F_2^{1,1}(x)$
at $\mathcal{Q}=15$~GeV \cite{Tanabashi:2018oca},
using the
Edin model (\ref{Edin}) and the normalization $N_p=0.426$.
This procedure can be repeated with the neutron structure function, giving
very similar results for the EMC effect for large nuclei. 
Smaller nuclei will be discussed in the next section.
}
\label{fig:F2Prot}
\end{figure}

The implications of EMS, which were given above, are intuitive in physical position space.
However, structure functions, relevant for DIS observables, are given
in terms of the parton momentum fraction $x$.
To be able to make  quantitative predictions
about the impact of EMS on DIS one needs a model which describes
this transition from position to ``x'' space.
For this purpose, we will utilize a description relying
on the structure of the Edin model proposed and described in \cite{Edin:1998dz,Edin:1999ep}.
In this model the dominant contribution to the $F_2^{2,1}$ structure function of 
Deuterium will be taken as a one to one mix of a proton and a neutron.
We define thus $F_2^{1,1}\equiv F_2^{2,1}/2$.
By using this definition, one neglects the possible EMS effects
of Deuterium. However, for Deuterium the EMS effects should be the smallest ones
available as one can see for example from figure \ref{fig:slopeQ}.
This structure function takes the form
\be\label{Edin}
F_2^{1,1}(\tilde \sigma, x)=N_p \exp\left( - \frac{x^2}{4 \tilde \sigma^2}\right) \cdot {\mbox{erf}}\left( \frac{1-x}{2 \tilde \sigma}\right),
\ee
where the dimensionless parameter $\tilde \sigma$ is given in terms of the proton mass $m_p$
and the proton (neutron) radius $r_p$ measured in the rest frame of the proton (neutron)
\be\label{sigma}
\tilde \sigma =\frac{1}{r_p m_p}=0.213.
\ee
Fitting only one parameter, namely the normalization $N_p$, one obtains a
 good fit  of the proton structure function, in the  $x$-range, which is relevant
for our study, as shown in figure \ref{fig:F2Prot}.
In the following subsections the model~(\ref{Edin}) will be used 
to estimate the EMS effects on a nucleus composed of deformed nucleons.
In a deformed nucleon, the radius becomes dependent on the direction
it is measured. In this case, $r_p$ will refer to the radial direction parallel
to the motion of the nucleon as seen from the center of mass frame of the collision
measuring this nucleon. The reason for this is that
 the fractional momentum $x$ is defined in the same direction.

\subsection{EMC fit with EMS}

In this subsection, the EMS idea will be implemented
in the Edin model  (\ref{Edin}). This will allow 
to fit EMC data from DIS with Iron nuclei. 
Iron is chosen because it has a very good
data quality for large momentum fraction~``x''.

The EMC effect is typically expressed in terms of 
a normalized ratio. Let us consider Iron ($A=56$ and $Z=26$)
and Deuteron ($A=2$ and $Z=1$)
\bea\label{ratio}
R(x)&=&\frac{F_2^{56,26}(x)}{28 \cdot F_2^{2,1}(x)}
\approx \frac{F_2^{56,26}(x)}{56 \cdot F_2^{1,1}(x)}.
\eea
Since at very low $x<0.2$, the EMC effect is dominated by the well understood
parton shadowing and the anti-shadowing of the pion cloud 
and since for very high $x>0.85$, 
correlations and multiple scatterings become dominant \cite{Hen:2016kwk}
and error bars become large,
the following analysis will
be restricted to the range $0.2< x < 0.85$.

\begin{figure}[h!]
\includegraphics[width=\columnwidth]{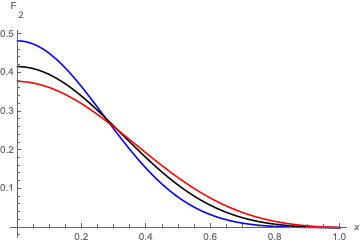}
\caption{
Deformed structure functions of the model (\ref{Edin}) using
the different longitudinal radii (\ref{ri})
for $\delta = 0.17$.
The blue curve represents $F_2^{1,1}\left(\tilde \sigma_a,x\right)$ for the stretched nucleons,
the black curve represents the undeformed $F_2^{1,1}(\tilde \sigma_b ,x)$,
and the red curve represents $F_2^{1,1}(\tilde\sigma_c,x)$ for the squeezed nucleons.
}
\label{fig:F2deformed}
\end{figure}

As a first quantitative approach to this idea, lets stick to
the regions shown in figure \ref{fig:WithStress} and assume
that each of them has nucleons with a constant deformation.
According to the EMS effect described above,
the iron nucleus is composed of a fraction of $p_a$ stretched, 
$p_b$ undeformed, and $p_c$ squeezed nucleons, giving a combined
structure function
\be\label{F256}
\frac{
F_2^{56,26}}{56}=  p_a F_2^{1,1}(\tilde \sigma_a, x)+
p_b F_2^{1,1}(\tilde \sigma_b, x)+
p_c F_2^{1,1}(\tilde \sigma_c, x).
\ee
The difference between the three contributions (a, b, c)
is that they have different  proton thickness in direction $\hat i$: $r_i$. 
According to (\ref{sigma})
this  corresponds
to different $\tilde \sigma_i$.
For  a comparable width and length deformation, this means
\bea \label{ri}
 r_a&\approx& r_p (1+\delta),\\ \nonumber
 r_b&\approx & r_p, \\ \nonumber
 r_c&\approx& r_p (1-\delta).
\eea
Please note that even though the EMS can deform
the radius $r_i$ it does not change the nucleon mass,
which will remain constant in equation (\ref{sigma}).
According to the model (\ref{Edin}), for each radius $r_i$ one expects a slightly deformed 
structure function $F_2^{1,1}$ as shown in
figure \ref{fig:F2deformed}.

\begin{figure}[h!]
\includegraphics[width=\columnwidth]{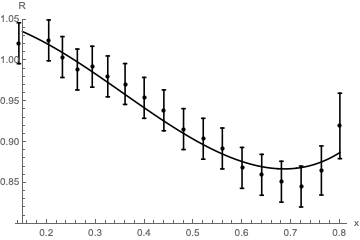}
\caption{Fit of the EMC effect for the iron nucleus 
given from a compilation from charged lepton DIS \cite{Bodek:1983qn,Bari:1985ga,Benvenuti:1987az,Gomez:1993ri,Dasu:1993vk,Schienbein:2009kk} 
with the EMS adapted Edin model with the probabilities (\ref{pabc}), and
for $\delta = 0.17$. For lower for $x<0.7$ the error bars were taken at an averaged value of $\pm 0.025$,
while for large $x$, larger errors are considered.
}
\label{fig:EMCFe}
\end{figure}

Those different structure functions can be combined
in (\ref{F256})
and subsequently used to form the EMC ratio (\ref{ratio}) which then reads
\be\label{ratioEMS2}
R(x)=\frac{p_a F_2^{1,1}(\tilde \sigma_a,x)+
p_b F_2^{1,1}(\tilde \sigma,x)+
p_c F_2^{1,1}(\tilde \sigma_c,x)}{F_2^{1,1}(\tilde \sigma,x)}.
\ee
This is the EMS model of the EMC effect.
The parameters are $p_a, \, p_b,\, p_c$, and $\delta$.
For scattering processes with long range interactions, one has further the constraint $p_a+p_b+p_c=1$.
Taking a relative deformation of $\delta =17 \%$ one can fit
the iron EMC data, as shown in figure \ref{fig:EMCFe}.
One realizes that the parameters
\bea\label{pabc}
p_a&=&0.50,\\ \nonumber
p_b&=&0.24, \\ \nonumber
p_c&=&0.26,
\eea
give a good fit of the EMC effect, not only
showing the downhill slope at $0.15<x<0.7$,
but also reproducing the inversion and depth at $x\approx 0.7$.
This fitting procedure was repeated, varying the parameters
$\delta, \, p_a,\, p_b, \, p_c$ subject to the constraint, and the numerical given 
values turned out to give the best result.
As a sanity check, one further realizes that the values in (\ref{pabc}) 
are pretty close to the values that arose from the illustration example in figure \ref{fig:WithStress}.
A sanity check for $\delta=\delta(Z)$ will be discussed in subsection \ref{subsec:MITbag}.

\subsection{Charge dependence}
\label{subsec:Charge dep}

\begin{figure}[h!]
\includegraphics[width=\columnwidth]{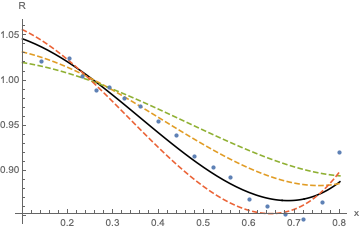}
\caption{
Ratio (\ref{ratio}) predicted for the same geometric distribution (\ref{pabc})
but with different  $\delta=\delta(Z)$.
The black curve is for $\delta = 0.17$, which was shown with the corresponding
data points already in figure \ref{fig:EMCFe}.
The red, yellow, and green dashed curves are for $\delta = 20 \%, 13\%, 10 \%$ respectively.
}.
\label{fig:varyp}
\end{figure}

\begin{figure}[h!]
\includegraphics[width=\columnwidth]{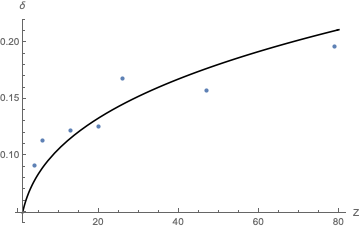}
\caption{
Deformations $\delta$ as a function of $Z$.
Here, the parameters $\delta(1), p_a, p_b, p_c$ are the ones
given above.
}.
\label{fig:deltaofZ}
\end{figure}

At first sight, the distribution (\ref{pabc}) should
be independent of the number of protons and neutrons
in a given nucleus as long as the ratio between proton and neutrons $Z \sim A$
does not vary too much.
The well-measured dependence on the atomic number $A$ enters indirectly
through the average deformation $\delta$.
The main increase of the deformation $\delta$ in (\ref{ri}) 
would be due to larger average electric fields $\bar E$ 
produced by a larger total charge $Z$ (\ref{barE})
which is proportional to the total number of nucleons
\be\label{proportionalities}
\delta \sim \bar E \sim Z^{1/3} \sim A^{1/3}.
\ee
Figure \ref{fig:varyp} shows how the ratio (\ref{ratioEMS2}) changes
for different choices of $\delta$.
One notes clearly that a larger $\delta$ produces 
a steeper ratio (\ref{ratio}), while a smaller $\delta$ produces a
flater ratio.
Plotting the $\delta=\delta(Z)$, necessary to 
fit the experimental results for the ratio (\ref{ratioEMS2}) for different nuclei 
($He,Be, C, Al,Ca,Fe,Ag,Au,$ all with fixed (\ref{pabc}))
one obtains a beautiful confirmation
 of the $\delta \sim Z^{(1/3)}$ dependence (\ref{proportionalities}), expected
 from the above model of the EMS effect.
 The best fit of the $\delta(Z)$ is obtained for
 \be\label{propdelta}
 \delta(Z) = 0.05 \cdot Z^{1/3},
 \ee
which is shown in figure \ref{fig:deltaofZ}.
This does not only confirm the linearity estimate of figure 
\ref{fig:slopeQ} with a concrete model, but
it also shows that both treatments give the same order of magnitude of deformation
through the proportionality factors of the deformation $\delta(1)=0.05$ in (\ref{propdelta})
and of $d'=0.07$ in (\ref{EMCmod2}).

\subsection{Scale dependence for $\mathcal{Q}^2>M_p^2$}
\label{subsec:Q2dep}

An essential feature of the EMC effect is that
the ratio (\ref{ratio}) is very robust under changes of  $\mathcal{Q}^2$  \cite{Eskola:1998df}.
It was, for example, observed that changes of the EMC slope when going from 
$\mathcal{Q}^2= 3 GeV^2$ to $\mathcal{Q}^2= 100 GeV^2$ are compatible with zero.

Since the above explanation of the EMC effect uses deformed structure functions,
and since those structure functions have a strong $\mathcal{Q}^2$-dependence,
it is at first puzzling, how $\mathcal{Q}^2$-robustness could arise from such a model.
However, as it will be shown below, it turns out that in practice,
the  $\mathcal{Q}^2$ cancels out almost entirely from the electromagnetic stress model.
There are two main reasons for this.
\begin{itemize}
\item[i)]
The $\mathcal{Q}^2$-dependence of the structure functions $F_2(x, \mathcal{Q}^2)$
is most important for $x<0.1$, while for larger $x>0.1$ there is only a few percent change
when going from $\mathcal{Q}^2= 3 GeV^2$ to $\mathcal{Q}^2= 100 GeV^2$ \cite{Lai:1999wy}.
Since the slope of the EMC effect only relies on $0.1<x<0.7$ data, the $\mathcal{Q}$-dependency
of this slope should also be reduced to this order of magnitude.
\item[ii)]
Within the simple implementation in the Edin model (\ref{ratioEMS2}), the EMC effect  $R(x)$ is a ratio
of deformed structure functions, all evaluated at the same fraction $x$.
At a given value of $x$, $\mathcal{Q}^2$ induced changes will apply almost multiplicatively
in both, the structure functions of the numerator and of the denominator. Those effects tend to cancel
each other, and the ratio $R(x)$ remains almost invariant under changes of $\mathcal{Q}^2$.
\end{itemize}

\begin{figure}[h!]
\includegraphics[width=\columnwidth]{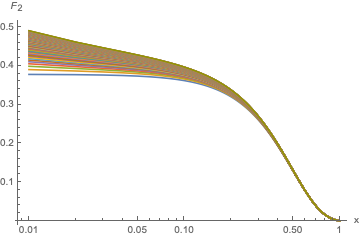}
\caption{
$\mathcal{Q}^2$-dependence of $F_2^{1,1}(\tilde \sigma_a,x)$. The lower curve
is for $\mathcal{Q}^2= \mathcal{Q}^2_0$ while the upper curve is for $\mathcal{Q}^2= 100 GeV^2$.
The evolution (\ref{DGLAPben}) was calculated with $\bar P_{q,g}$, but the other
splitting functions give very similar results.
}
\label{fig:F2bundle}
\end{figure}

In order to exemplify this argument quantitatively, one can consider the ratio (\ref{ratioEMS2}),
with the parameters (\ref{pabc}) at an initial scale $\mathcal{Q}_0^2\approx k M_p^2$.
The $\mathcal{Q}^2$-dependence of each of the structure functions is given by the 
DGLAP evolution equations \cite{Altarelli:1977zs,Lai:1999wy}  
\be\label{DGLAPben}
\frac{d}{dt} F_2(x,t)=\frac{\alpha_s(\mathcal{Q}^2)}{2 \pi }\int_x^1 dy \frac{F_2(y,t)}{y}\cdot \bar P\left( \frac{x}{y}\right)
+\mathcal{O}\left( \alpha_s^2\right),
\ee
where $t=\mathcal{Q}^2/\mathcal{Q}_0^2$, and where $\bar P$ is a weighted average of the splitting functions
$P_{q,q}, \, P_{q,g}, \, P_{g,q}, \, P_{g,g}$, whose detailed form does not affect the following discussion.
By integrating (\ref{DGLAPben}), one obtains a structure function at large $\mathcal{Q}^2$ from
an initial structure function at $\mathcal{Q}_0^2\approx k M_p^2$.
Figure \ref{fig:F2bundle} shows the deformation of the structure function
$F_2^{1,1}(\tilde \sigma_a,x,\mathcal{Q}^2)$ when going
from $\mathcal{Q}^2= \mathcal{Q}^2_0$ to $\mathcal{Q}^2= 100 GeV^2$.
One notes that changes are much stronger for $x<0.1$ as anticipated in $i)$,
but still at $x\approx 0.1$ a ten percent change is perceivable. This shows that the argument
$i)$ is insufficient to explain the observed robustness of $R(x)$ under substantial changes in $\mathcal{Q}^2$.

\begin{figure}[h!]
\includegraphics[width=\columnwidth]{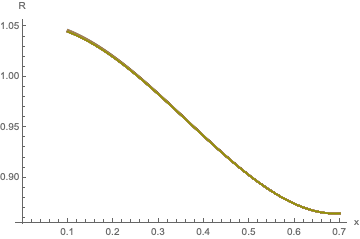}
\caption{
$\mathcal{Q}^2$-dependence of $R(x)$.}
\label{fig:Rbundle}
\end{figure}

\begin{figure}[h!]
\includegraphics[width=\columnwidth]{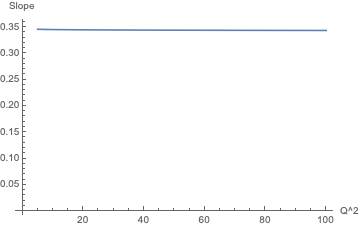}
\caption{
$\mathcal{Q}^2$-dependence of the $R(x)$ slope between $0.1<x$ and $x<0.7$.}
\label{fig:SlopeofQ2}
\end{figure}

One can evolve all the structure functions in (\ref{ratioEMS2}) and thus calculate
the ratio $R=R(x,\mathcal{Q}^2)$. The result of this procedure is shown in figure \ref{fig:Rbundle}.
One notes that the different structure functions of figure \ref{fig:F2bundle},
are now compressed to a single, almost indistinguishable, bundle of curves in figure \ref{fig:Rbundle}. 
This compression is
a result of the cancellation effect, anticipated in argument $ii)$
and it  translates into a $\mathcal{Q}^2$-independence of the ratio $R$.
Fitting the slope of the curves in figure \ref{fig:Rbundle} between $0.1<x$ and $x<0.7$ gives the slope
as a function of $\mathcal{Q}^2$. The result of this procedure is shown in figure
\ref{fig:SlopeofQ2}.
Clearly the $\mathcal{Q}^2$-dependence of 
the EMC slope turns out to be negligible, due to the cancellation effect 
described in ii).

\subsection{Scale dependence for $\mathcal{Q}^2<M_p^2$}
\label{subsec:Q2dep2}

In the introduction, the 
theoretical estimate (\ref{delta1}) for the relative deformation $\delta$ of a nucleon 
in a tiny nucleus was given to be of the order of $10^{-3}$.
This seems to be in tension with the 
 $\delta(1)\approx0.05$ we found from explaining the EMC effect with EMS deformations (\ref{propdelta}).
In order to understand this difference, one has to keep two effects in mind.
\begin{itemize}
\item
Basically, all nucleon properties are affected by the surrounding nuclear medium.
For the case of polarizability, this means that strong and mesonic interactions
with the surrounding nuclear matter, can ``soften the skin'' of an individual nucleon.
This would make the nucleon much more susceptive to electromagnetic stress.
This enhancement might explain the relatively large deformations in the EMS-EMC explanation (\ref{propdelta}).
\item
Both $\alpha_s$ and $\alpha_E$ are $\mathcal{Q}^2$-dependent quantities. 
For $\mathcal{Q}^2>M_p^2$, it was shown above that there is a cancellation of $\mathcal{Q}^2$ effects
when calculating $R(x)$.
However, for smaller $\mathcal{Q}^2$ and for the polarizability underlying the estimate (\ref{delta1})
there is no reason to expect such a cancellation.
For example,
in \cite{Holstein:2013kia} the generalized spin-independent polarizability of a proton was given as
\be\label{alphaQ}
\frac{\alpha_E(\mathcal{Q}^2)}{\alpha_E(0)}= \left(1- \frac{7\mathcal{Q}^2}{50 M_\pi^2}+\frac{81\mathcal{Q}^2}{2100 M_\pi^4}+ \mathcal{O} \left(\frac{\mathcal{Q}^6}{M_\pi^6}\right) \right).
\ee
The estimate (\ref{delta1}) was derived from measurements at $\mathcal{Q}^2\ll M_\pi^2$,
while the EMC effect is measured at  $\mathcal{Q}^2\gg M_\pi^2$.
Even though the approximation underlying the calculation of (\ref{alphaQ})
collapses for larger $\mathcal{Q}$,
it is clear  from (\ref{alphaQ}) that a relative increase
of $\delta=\delta(Q^2)$ by a factor of 50 is certainly possible when going from $\mathcal{Q}^2\ll M_\pi^2$
to $\mathcal{Q}^2 \approx M_\pi^2$.
\end{itemize}

\subsection{Deformation estimate in a bag model}
\label{subsec:MITbag}

The preceding sections give on the one hand an estimate of
the deformation $\delta$ between two neighboring protons
at very small ${\mathcal{Q}}^2$ (\ref{delta1})
and on the other hand the amount of deformation needed to explain the EMC result 
at larger ${\mathcal{Q}}^2$~(\ref{propdelta}).
In this subsection, a simple order of magnitude estimate will be given,
which will show that both results are compatible.

The nucleon is held together by the strong interaction
with the coupling constant $\alpha_s$.
This confining interaction can be expected to resist deformations.
The EMS effects, in contrast, are caused by electromagnetic interactions with the coupling
constant $\alpha_e$. If this coupling
would be zero or extremely small, even the strongest external electric field could
not deform a nucleus.\\
Thus, one can expect that the dimensionless deformability $\delta$
increases with growing $\alpha_e$, but decreases with growing $\alpha_s$.
This suggests at leading order in $\alpha_e$ and for comparable electric and strong
charge number $Z_e\approx Z_s\approx 1$
\be
\delta\sim \frac{\alpha_e}{\alpha_s}.
\ee
However, both couplings are scale-dependent, in particular, $\alpha_s=\alpha_s(\mathcal{Q}^2)$ can change significantly
with $\mathcal{Q}$.
Therefore, the ratio $\delta$ inherits the $Q^2$ dependence from the couplings
\be\label{deltaQ}
\delta(\mathcal{Q}^2)\sim \frac{\alpha_{e}(\mathcal{Q}^2)}{\alpha_s(\mathcal{Q}^2)}.
\ee

One can also deduce (\ref{deltaQ}) from a
more fundamental model. 
A good candidate for this procedure is
the MIT bag model.
The input of this model is the shape and strength
of the confining strong potential (which is proportional to $\alpha_s$). 
In order 
to do this calculation directly, one needs, however, the analytic form of 
the bound state solutions.
There are few analytical solutions for quantum mechanical potentials known.
For simplicity, the harmonic oscillator in three dimensions will be used.
Given the distance between two nuclei $\bar r$ and the electric polarizability 
$\alpha_E$ one can
estimate the deformation due to a charged neighbor (see \ref{delta1})
\be
\delta\approx \frac{\alpha_E}{\bar r^3}.
\ee
For a given model of the nucleus, the polarizability can be calculated from~\cite{Hecking:1981fsu}
\be\label{sum}
\alpha_E=2 \alpha_e  \sum_{\lambda} \frac{|\langle \psi_\lambda |z|\psi_0 \rangle |^2}{E_\lambda-E_0},
\ee
where $\psi_\lambda$ are the excited eigenfunctions of a given quantum mechanical potential,
which will depend on $\alpha_s$.
For the purpose of finding a ratio of couplings, like the one in (\ref{deltaQ}), one has to keep
track of the  coupling constants when computing (\ref{sum}). 
One can exemplify this for the 3-D harmonic oscillator with the potential
\be\label{potHO3}
V(\vec x)=\frac{1}{2} \mu \alpha_s \vec x^2.
\ee
The eigenfunctions  $\psi_{k,l,m}$ of this problem are products of spherical harmonics
and generalized Laguerre polynomials~\cite{Messiah:1979eg}.
With these, the non-vanishing integrals of the numerator of (\ref{sum}) behave as
\be\label{prop1}
\langle \psi_\lambda |z|\psi_0 \rangle\sim \alpha_s^{-1/4}.
\ee
The energy differences in the denominator behave as
\be\label{prop2}
E_\lambda-E_0\sim \sqrt{\alpha_s}.
\ee
Inserting (\ref{prop1} and \ref{prop2}) in (\ref{sum}) one finds that the polarizability
scales with the couplings as
\be
\label{sum2}
\alpha_E\sim \frac{\alpha_e}{\alpha_s}.
\ee
Next, one takes a fixed distance between two nuclei (meaning that $\bar r$ is not scaling with ${\mathcal{Q}}^2$).
This is a reasonable assumption since in the suggested model, the deep inelastic scattering simply probes
the internal structure of the individual nucleons.
One finds
\be\label{deltaQderived}
\delta(Z=1)Ê\sim \frac{\alpha_e}{\alpha_s}.
\ee
Thus, the ad-hoc relation (\ref{deltaQ}) can be justified from
a more fundamental point of view.
 
Let's shortly explore the consistency of this relation at different scales
\begin{itemize}
\item $(\mathcal{Q}^2\rightarrow \infty)$:\\
For very high energy processes $(\mathcal{Q}^2\rightarrow \infty)$, 
it is well known that 
$\alpha_s(\mathcal{Q}^2)\rightarrow 0$ and thus $\delta \rightarrow \infty$.
This means that small external electromagnetic fields can 
produce large changes in the shape of the nucleus.
The nucleon is extremely ``soft'', meaning that the constituents are not confined anymore.
This effect is famously known as asymptotic freedom \cite{Gross:1973id,Politzer:1973fx}.\\
Thus, the ratio (\ref{deltaQ}) reflects our intuitive understanding of confinement
and strong interactions at large ${\mathcal{Q}}^2$.
\item $(\mathcal{Q}^2\rightarrow \Lambda_{QCD})$:\\
If to the contrary, the strong coupling dominates the electromagnetic coupling,
all  corrections to the form of the nucleon should get suppressed.
At very small  $\mathcal{Q}$, close to the energy scale of QCD $\Lambda_{QCD}$, one expects 
$\alpha_e(0)\approx \frac{1}{137}$ and $\alpha_s(0)\approx \pi$~\cite{Deur:2005rp}.
Thus, the deformation gives
\be\label{delta00}
\delta(\Lambda_{QCD}))\approx\frac{1}{\pi^{} \cdot 137}\approx 0.002,
\ee
as an order of magnitude estimate.
This result has to be compared to the $\delta$ obtained from the 
low energy polarizability (\ref{delta1}) which only differs by a factor of two. 
This is a compelling agreement for this type of estimate.
\item Intermediate $(\mathcal{Q}^2)$:\\
The EMC effect is measured at intermediate scales energy
scales of about $3~GeV^2<{\mathcal{Q}}^2<100~GeV^2$.
In order to estimate the ${\mathcal{Q}}^2$ dependence of (\ref{deltaQ})
one can solve the beta function of QCD $\beta_s=-\frac{7 \alpha_s^2}{2\pi}$
and QED $\beta_e=\frac{2\alpha_e^2}{3 \pi}$ subject to the initial conditions
 $\alpha_s(\Lambda_{QCD})=\pi$ and $\alpha_e(\Lambda_{QCD})=1/137$ 
giving
\be
\delta ({\mathcal{Q}})\approx\frac{3}{2^{}}\frac{(2+7 \ln k)^{} }{411 \pi-2\ln k},
\ee
where $\ln k = \ln ({\mathcal{Q}}/\Lambda_{QCD})$.
This gives for the aforementioned energy range
\be\label{deltaQint}
\delta \approx 0.02\dots 0.04.
\ee
\end{itemize}
Relation (\ref{deltaQint}) is the order of magnitude estimate 
for $\delta$ in the context of the EMC experiment.
This quantity should be compared to $\delta\approx 0.05$ (see relation 3.9), and $d'\approx 0.07$ (see Fig.2)
obtained from assuming that the EMC effect can be explained
from the hypothesis of an EMS mechanism.
Clearly, these quantities lie closely in the same ballpark, 
justifying the consideration
of the EMS mechanism in the first place.

\section{Small atomic nuclei}\label{sec:sum}

For small atomic nuclei, the addition of every single nucleon will change the geometry of
the nucleus. Thus,
 the homogeneity assumptions of the previous two sections
are doomed to fail. This can be seen already from the fact that 
for $A=N+Z=2$, the EMC slope must be zero by construction.
The model (\ref{EMCmod2}) does not have this feature
because it is constructed for large nuclei.
If one tries to include this initial condition in an analogous model 
\be \label{sconti}
s(A)=\left.\frac{d}{dx}
\frac{2\sigma^{A}(x)}{A\cdot \sigma^{2,1}(x)}\right|_{x_{lin}}
 =d' \cdot (A-2)^{1/3},
\ee
one realizes that, for small nuclei, this model is far off from the data \cite{Seely:2009gt}.
A simple reason for this failure is that, the estimate for the average electric field (\ref{barE}) is not applicable
to small nuclei. In the following subsection, a discrete model for calculating this
average electric field is presented.

\subsection{A simple discrete model}
\begin{figure}[h!]
\includegraphics[width=\columnwidth]{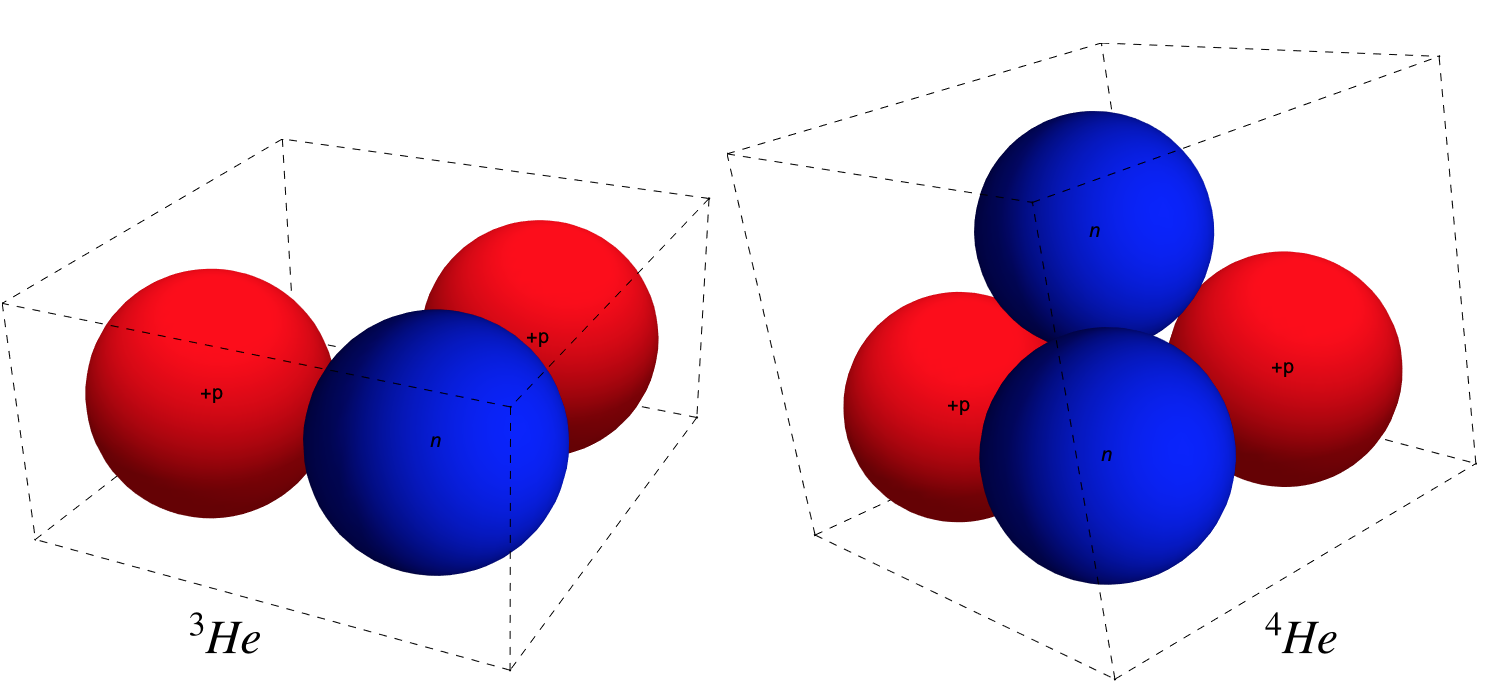}
\caption{}
\label{fig:He34}
Geometric configuration of the $^3He$, $^4He$ nuclei.
The blue spheres indicate the relative positions of neutrons and
the red spheres indicate the relative positions of protons.
For simplicity, the opening angle of the triangle $p-n-p$ was chosen to be $\pi/2$. 
\end{figure}

\begin{figure}[h!]
\includegraphics[width=\columnwidth]{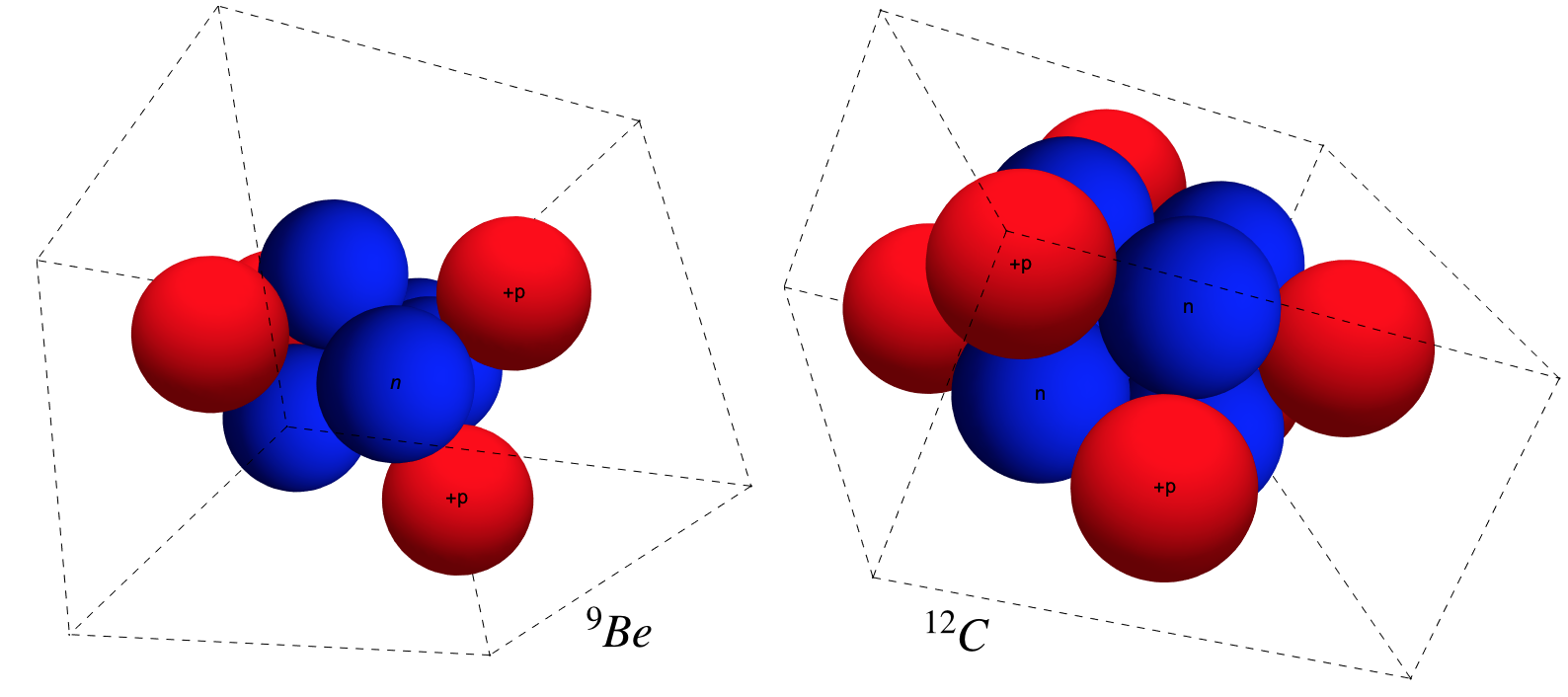}
\caption{}
\label{fig:Be9C12}
Geometric configuration of the $^9Be$, $^{12}C$ nuclei.
The blue spheres indicate the relative positions of neutrons and
the red spheres indicate the relative positions of protons.
For simplicity, the opening angle of the triangle $p-n-p$ was chosen to be $\pi/2$. 
\end{figure}

Given the difficulties of the continuous model, when it comes to small nuclei, 
one can try to describe these nuclei with the polar opposite simplification, a rigid discrete model.
In this model,   protons $p$ are assumed to be   point charges inside of a sphere of radius $r_p$, while
neutrons $n$ are assumed to be neutral spheres of approximately the same radius. 
In stable nuclei $p-n$  and  $n-n$ are bound together, putting their spheres in contact,
while $p-p$ experience an interplay between atomic attraction and electromagnetic repulsion,
resulting in a distance $d>2r_p$ between the centers both nuclei.
Following these simple construction rules, one can build a rigid discrete charge distribution for 
the light isotopes $^2D$, $^3He$, $^4He$, $^9Be$, and $^{12}C$, whose EMC slope is  given in \cite{Seely:2009gt}.
For $^2D$ the geometric configuration is trivial, and for $^3He$ and $^4He$ the configuration
is shown in figure \ref{fig:He34}.
One realizes that 
the best configurations (according to the above rules) for $^9Be$ and $^{12}C$
are the ones given in  figure \ref{fig:Be9C12}.
One notes that  $^9Be$ is composed of two $^4He$
and $^{12}C$ is made of three $^4He$.
In both cases electric repulsion lets the electrically charged protons 
accommodate oriented outwards.  The $^9Be$ nucleus contains one additional $n$,
which breaks the discrete symmetry of $^8Be$, by aligning outside of this configuration.

\subsection{EMS and EMC  for small nuclei}

Given the geometric configurations \ref{fig:He34}, or \ref{fig:Be9C12}, it
is straight forward to calculate the external electrical  field $\vec E_i $ at the  nucleon position $\vec x_i$,
produced by all the surrounding protons
\be\label{Ebardis}
\vec E_i(\vec x_i)= \sum_{\vec x_j, j\neq i}^{\tiny{\text{protons}}} \frac{q_j}{4 \pi \epsilon_0} \frac{\vec x_j- \vec x_i}{|\vec x_j- \vec x_i|^3}.
\ee
The average of the absolute values of these external
electromagnetic fields is
\be
\bar E(A,Z)=\frac{1}{A}\sum_{i=1,...A} |\vec E_i(\vec x_i)|.
\ee
This field can now be taken as source for the electromagnetic stress (\ref{EMCmod})
acting in the nucleons of small nuclei.
In analogy to (\ref{EMCmod2}) one expects the EMC slope $s$
to be
\be\label{EMCdiscrete}
s(A,Z)=\left.\frac{d}{dx}
\frac{2\sigma^{A,Z}(x)}{A\cdot \sigma^{2,1}(x)}\right|_{x_{lin}}
\approx  d''\left(\bar E(A,Z)-\bar E(2,1)\right),
\ee
where  $\bar E$ is calculated from (\ref{Ebardis}).
Here, the average electric field $\bar E(2,1)$  has to be subtracted,
since the EMC ratio is measured with respect to~$^2D$.
Just like in the continuous model (\ref{EMCmod2}), the universal proportionality
constant $d''$ has to be determined by fitting the data.

\begin{figure}[h!]
\includegraphics[width=\columnwidth]{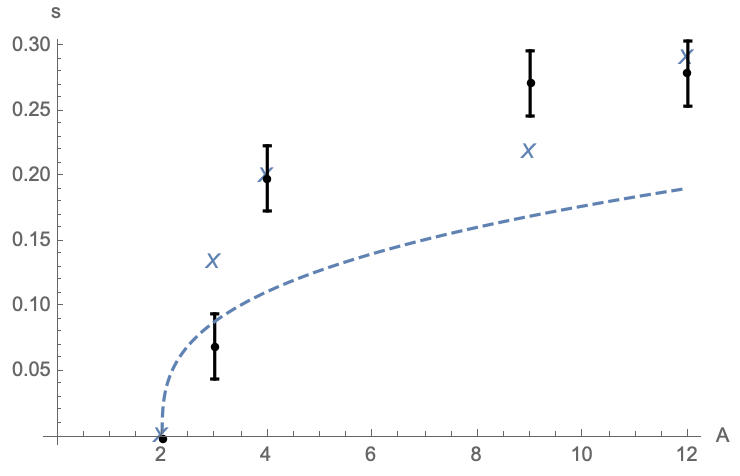}
\caption{Black: EMC slope measured for $^3He$, $^4He$, $^9Be$, and $^{12}C$ \cite{Seely:2009gt}.
Dashed line: Continuous model (\ref{sconti}) fitting the data for $A$ up to 160.
Blue $X$: Discrete model (\ref{EMCdiscrete}) with  the best fit value $d'' = 0.44$ and the electric field
from the configurations shown in figures \ref{fig:He34} and \ref{fig:Be9C12}.}
\label{fig:EMCsmall}
\end{figure}

In figure \ref{fig:EMCsmall}, the result of (\ref{EMCdiscrete})
is compared to the fit using the continuous model (\ref{sconti}) and to the
data points given in \cite{Seely:2009gt}.

One notes that the dashed line of the  continuous model (\ref{sconti}) is not able to fit the data in this range, while
the discrete EMS model (\ref{EMCdiscrete}) is doing a better job of fitting the EMC 
data for small nuclei. 

However, one has to keep in mind that the continuous model
is, by construction, meant to work for large nuclei, while the discrete model has not been
tested for nuclei larger than $A=12$. The construction of these larger nuclei
is left for future work.
Further effects, which will be considered in future investigations, are
those related to additional degrees of freedom, such as magnetic fields, rotation, or vibration
(note, that even the nuclear shell model has problems with explaining observed nuclear densities
when applied to smaller nuclei \cite{Nortershauser:2008vp,Krieger:2012jx}).
The purpose of this section was not to come up with such a model, but to show that
the EMS effect has a good potential of explaining the local nature of the EMC effect, which becomes prominent
when one deals with small atomic nuclei.

\section{Summary and Outlook}\label{sec:sum}

\subsection{Summary}
This article explored the  impact
of electromagnetic stress on the nucleon structure.
As a possible manifestation of this effect,  
the dependence of nuclear structure functions on external 
electromagnetic forces was studied.
Simple and straight forward assumptions lead
to a remarkably good description of the charge dependence
of the EMC effect.
This was first shown assuming a linear stress dependence
of the nuclear structure functions (\ref{EMCmod2}) and then confirmed
in an explicit model (\ref{ratioEMS2}).
The explicit model further captures the minimum and the rise of the EMC effect
at larger $x \approx 0.7$.
In subsection \ref{subsec:Q2dep2} it is shown how the $\mathcal{Q}^2$-independence of the EMC effect arises naturally
within this model for $\mathcal{Q}^2>M_p^2$.
Finally, it is noted that the continuous models have to fail when applied to
small nuclei and that this can be fixed by studying the electromagnetic stress in
a more appropriate approximation. For example,  as shown in the previous section, 
even a straight forward discrete, and in many ways naive implementation
gives promising results with this respect.

The strength of the  EMS idea is
that it uses a very well known effect in nuclear physics, 
namely polarizability,
in order to give a description for different aspects of the EMC effect.
This explanation is, of course, not exclusive.
Different effects can add up and influence each other.
For example, in \cite{Weiner:1985ih},
it was shown that pionic effects could contribute significantly to
the polarizability of nuclei. Thus, one should also  expect 
an important modification of the polarization
from the pion excess~\cite{Ericson:1983um,Wiringa:1984tg,Berger:1984na,Berger:1987er}, 
which is one of the suggested explanations of the EMC effect.
Thus, it is probably
a long way to go until the true origin of the EMC effect is known with certainty.
Still, the EMS explanation is worth to be considered. It is actually very conservative
since it does
not need to invoke new intermediate states or otherwise unobserved effects.
\subsection{Future directions}
\label{subsec:future}

An interesting test which will be left for future investigation is that
one can calculate the EMS effect with other nucleus and nucleon 
 models~\cite{Baranger:1960qge,Jaffe:1978bu,Brown:1979ui,Brodsky:1982nx,Brodsky:2000ii,Gutsche:2016gcd,DiPiazza:2011tq},
 which are different from the continuous fit (\ref{EMCmod2}), the
 Edin nucleon model with the ``three regions nucleus'' (\ref{ratioEMS2}), and the discrete model form small  
nuclei (\ref{EMCdiscrete}) used here.
In particular, for smaller nuclei, short-range correlations revealed important information \cite{CiofiDegliAtti:1989eg,Weinstein:2010rt,Fomin:2011ng,Arrington:2012ax,Hen:2012fm,Frankfurt:2012qs,Hen:2013oha},
which points towards local effects that go beyond a mean-field explanation.
The first step into this direction will be to generalize the approach (\ref{ratioEMS2}), which was used for continuous
distributions with probability fractions $p_a,\, p_b,\, p_c$. In this generalization, the probabilities would
be replaced by a sum of structure functions, one for each individual nucleon.
Since our model is also based on local modifications of structure functions, it would be very interesting
to explore such correlations.
Once a more realistic and more precise model is implemented, the systematic error bars of this study will reduce.
Thus one has to pay more attention to
the experimental error bars and their interpretation. In particular, considering the isoscalar corrections of the
SLAC data will then be important.

However, these and other improvements go beyond the scope of this paper.

\subsection{Acknowledgements}
Thanks to Marcelo Loewe Lobo and Matt Sievert for valuable comments
and suggestions and 
 Ina Gruber for thorough proofreading.
Also thanks to Giorgio Torrieri for critical remarks.
This work was supported by Fondecyt 1181694.

\section*{Appendix}

The picture behind relation (\ref{delta1}) is Figure \ref{fig:NoStress} 
where one compares one of the nucleons
on the left-hand side with one of the deformed nucleons on the right-hand side. 
The change in radius $\Delta r$ of one of
the nucleons defining
\be\label{delta0}
\delta\approx \frac{\Delta r}{r}.
\ee
Since the extremes of the deformed nucleon 
are relatively small regions, they can
be assumed to good approximation to be homogenously charged.
Thus, $\Delta r$ is proportional to the amount of displaced charge $\Delta q$, 
which implies
\be\label{delta1a}
\delta \approx \frac{\Delta q}{q}.
\ee
The dipole moment of the deformed nucleon is $p =r \Delta q $.
The same dipole moment is also given in terms of the polarizability $\alpha_E$  through 
$p\sim \alpha_E |\vec E|$~\cite{Holstein:2013kia}, where the electric field of the neighboring nucleon
is $\vec E= \frac{q}{r^2}$. Inserting these relations into (\ref{delta1a})
one arrives at
\be\label{deform}
\delta \sim  \frac{\alpha_E}{r^3},
\ee
which is the relation used in (\ref{delta1})


\end{document}